# Spin Manipulation by Giant Valley-Zeeman Spin-Orbit Field in Atom-Thick WSe$_2$

Xinhe Wang[1,†], Wei Yang[1,†], Wang Yang[2,†], Yuan Cao[1,†], Xiaoyang Lin[1,*], Guodong Wei[1], Haichang Lu[1], Peizhe Tang[3], and Weisheng Zhao[1,*]

[1] *Fert Beijing Institute, MIIT Key Laboratory of Spintronics, School of Integrated Circuit Science and Engineering, Beihang University, Beijing 100191, China*
2 *Department of Physics and Astronomy and Stewart Blusson Quantum Matter Institute, University of British Columbia, Vancouver, B.C., V6T 1Z1, Canada*
3 *School of Materials Science and Engineering, Beihang University, Beijing ,100191, China*
E-mails: *XYLin@buaa.edu.cn* (X.L.); *weisheng.zhao@buaa.edu.cn* (W.Z.)

The phenomenon originating from spin-orbit coupling (SOC) provides energy-efficient strategies for spin manipulation and device applications. The broken inversion symmetry interface and resulting electric field induce a Rashba-type spin-orbit field (SOF), which has been demonstrated to generate spin-orbit torque for data storage applications. In this study, we found that spin flipping can be achieved by the valley-Zeeman SOF in monolayer WSe$_2$ at room temperature, which manifests as a negative magnetoresistance in the vertical spin valve. Quantum transmission calculations based on an effective model near the K valley of WSe$_2$ confirm the precessional spin transport of carriers under the giant SOF, which is estimated to be 650 T. In particular, the valley-Zeeman SOF-induced spin dynamics was demonstrated to be tunable with the layer number and stacking phase of WSe$_2$ as well as the gate voltage, which provides a novel strategy for spin manipulation and can benefit the development of ultralow-power spintronic devices.

## I. INTRODUCTION

The relativistic interaction between the spin and momentum degree of a traveling electron is known as spin-orbit coupling (SOC), which induces an effective magnetic field called spin-orbit field (SOF) in the frame of motion[1, 2]. The SOF in a structure with broken symmetry is a versatile way to manipulate spin without an external magnetic field[3, 4], and it induces various spin-orbit torques (SOTs) to switch the magnetization efficiently for data storage applications[5-7]. The spin dynamics directly driven by the SOF also promises ultralow-power logic applications. For example, the spin field-effect transistor (spin-FET)[8] utilizes a Rashba-type SOF at the interface, breaking the inversion symmetry to drive the spin flipping during diffusion[9]. However, because the effective field is generally no more than several Tesla[10-12], spin flipping requires a channel length exceeding a few microns. The spin decoherence within the micron-channel limits the operating temperature and prevents high-density integration of spin-FETs.

Compared to the Rashba-type SOF at the heterostructure interface, the broken symmetry of spatial inversion in two-dimensional transition metal dichalcogenides (2D-TMDs) induces a giant SOF up to hundreds of Tesla, which has opposite signs at the K and −K valleys to maintain time reversal symmetry, known as valley-Zeeman effect[13-18]. This SOF is Zeeman-type, that is, it is out-of-plane and not dependent on the direction of electron motion, as shown in FIG. 1(a), which can result in physical effects such as the spin-valley Hall effect, opto-valleytronics, the Ising superconducting state, and anisotropic spin relaxation[19-23]. Furthermore, as shown in FIG. 1(b), by altering the stacking phase to break or preserve the inversion symmetry in few-layer 2D-TMDs, the valley-Zeeman SOF can be turned on or off[24].

In this work, we reveal that the giant valley-Zeeman SOF can inspire an efficient spin flipping mechanism for high-density device integration at room temperature. When the carriers are spin-polarized in the plane, that is, perpendicular to the direction of the valley-Zeeman SOF, the giant valley-Zeeman SOF can result in a large deflection and even flipping of the in-plane spin. The spin dynamics can be detected through the vertical spin valve, where the spin polarized carriers are incident from the upper ferromagnetic (FM) layer and propagate vertically through the atom-thick TMDs, as shown in FIG. 1(c). Generally, the magnetoresistance (MR) of vertical spin valve is defined as MR=$(R_{AP}-R_P)/R_P$, where $R_P$ and $R_{AP}$ are the resistances when the FM electrodes are in parallel (P) and anti-parallel (AP) states respectively. In the presence of valley-Zeeman SOF, the MR actually depends on the relative relations between the spin polarization of the transmitted carriers and the magnetization of the bottom FM electrode, and thus contains the information of spin dynamics. Although the two valleys have opposite valley-Zeeman SOFs, the spin projections of the transmitted carriers on the magnetization of the FM electrode are always the same owing to the time reversal symmetry and thus contribute identically to the MR signal.

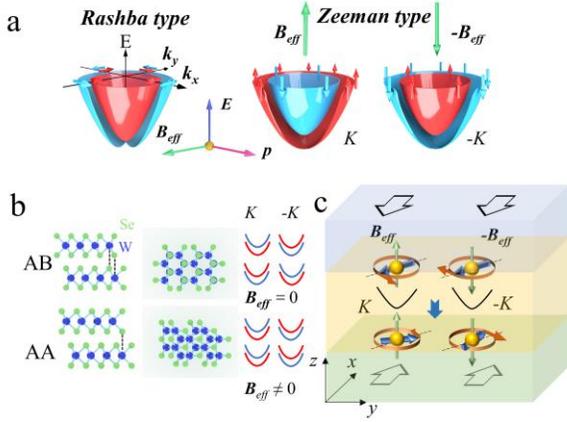

**FIG. 1.** Diagram of valley-Zeeman SOF and spin dynamics in $WSe_2$. (a) The distinct properties of Rashba-type and Zeeman-type SOF in 2D-TMDs. The red and blue paraboloids with arrows represent the spin sub-bands and corresponding spins. (b) Stacking dependence of valley-Zeeman SOF in bilayer $WSe_2$, where the blue and red lines represent the spin-up and spin-down sub-bands respectively, and the valley-Zeeman SOF exists only in AA stacking with broken inversion symmetry. (c) The vertical spin valve to manipulate and detect spin dynamics under giant valley-Zeeman SOF, where a 2D-TMD is sandwiched by two FM layers, and the top and bottom FM layers are used to generate and detect in-plane spins, respectively. When the carriers propagate vertically through the 2D-TMD, the spins are deflected and even flipped under the valley-Zeeman SOF.

In view of this physical scenario, we chose a $WSe_2$-based vertical spin valve to study the valley-Zeeman SOF-induced spin dynamics, where $WSe_2$ has been reported as a 2D-TMD with large spin splitting at valleys[17, 25]. We observed an MR oscillation between negative and positive values depending on the number of $WSe_2$ layers. In addition, the amplitude of the MR can be significantly regulated by changing the stacking phases of $WSe_2$. By comparing experiments with theoretical calculations in a variety of cases, we confirmed that these phenomena originate from the valley-Zeeman SOF-induced flipping of in-plane spin within atom-thick $WSe_2$. Our findings reveal a room-temperature spin manipulation strategy utilizing the giant valley-Zeeman SOF, which highlight the great potential of atom-thick semiconductor for spintronics and pave the way for the high-density integration.

To measure the valley-Zeeman SOF-induced spin dynamics through atom-thick $WSe_2$, we fabricated the vertical spin valve with different layers of $WSe_2$, as shown in FIG. 2(a). The bottom NiFe electrodes were prepared on the $SiO_2$(300nm)/Si substrate by photolithography, electron-beam evaporation and lift off method. A thin (~2nm) capping layer of gold was in-situ evaporated over the bottom NiFe to prevent oxidation, though the control experiments show that a thin layer of oxide doesn't make a significant difference to the results (Fig. S5(a)). To fabricate device with different layer number and stacking phase $WSe_2$, the single crystal flakes of $WSe_2$ were grown on $SiO_2$(300nm)/Si substrate by chemical vapor deposition (CVD), of which the layer number and stacking phase (AB or AA) were selected by optical microscope and further confirmed by photoluminescence (PL) and Raman spectra (see S1 for details). Then the natural $WSe_2$ crystal was transferred onto the bottom NiFe electrode by polyvinyl alcohol (PVA)-assisted ultraclean transfer method to obtain a cleaner interface [26]. Subsequently, Co(30nm) FM top electrode capped with Au(30nm) was patterned via electron-beam lithography and electron-beam evaporation. To rule out the interfacial effects, the device was annealed at 600 K for 2 h in the ultra-high vacuum (See S2 for details).

As shown in FIG. 2(b), the calculated band structure of monolayer $WSe_2$ indicates the existence of a large Zeeman splitting $2\Delta_{so} \approx 440$ meV at the valence band maximum (VBM), which is consistent with experiment results[27, 28]. If the Fermi level is close to the VBM, since the Fermi surfaces only locate in the vicinities of K and -K points, the transport properties and spin dynamics will be dominated by the K and −K valleys. Accordingly, to pin the Fermi level around the VBM of $WSe_2$ in the experiment, we used NiFe and Co as the bottom and top FM electrodes of the vertical spin valves. According to the band alignment in inset of FIG. 2(b), the Fermi level is located at $E-E_{VBM} \approx -0.5$ eV of $WSe_2$ in our devices. The band alignment was estimated based on work functions of Co (-5.0 eV), NiFe (-4.8 eV) and $WSe_2$ (-4.4 eV)[29] (see S4 for discussion of band alignment). The estimation about the location of Fermi level was supported by the metallic behavior of junction resistance (Fig.S2 and S3 for discussion).

The vertical transport measurements were performed by a four-terminal scheme, while an in-plane external magnetic field ($H$) was applied at 45° relative to the easy-axis of both FM electrodes, as the left inset of FIG. 2(a) shows. The directly measured results is composed of the anisotropic magnetoresistance (AMR) of FM electrodes and the MR of $WSe_2$ vertical spin valve. The MR of $WSe_2$, as shown in FIG. 2(c), were extracted by deducting the decomposed AMR[30]. We observed a definite negative MR as $R_{AP} < R_P$, which was robust even at room temperature (FIG. 2(d)). The amplitude of the negative MR ratio decreases with temperature, which can be fully described by Julliere's model $|MR|= 2P_1P_2/(1-P_1P_2)$ considering the thermal depolarization of the magnetism $P(T)=P_0(1-\alpha T^{3/2})$. The factor α is fitted to be $8.68\times10^{-5}$ K$^{-3/2}$ and is consistent with the value reported in literature[31-34] (see S5 for fitting details). Additionally, various bias currents have no significant effect on the MR ratio (FIG. 2(e)), which is different from previous reports of 2D-TMDs vertical spin valves[32, 33] and magnetic tunnel junction devices, indicating a physical mechanism that is different from that of these devices.

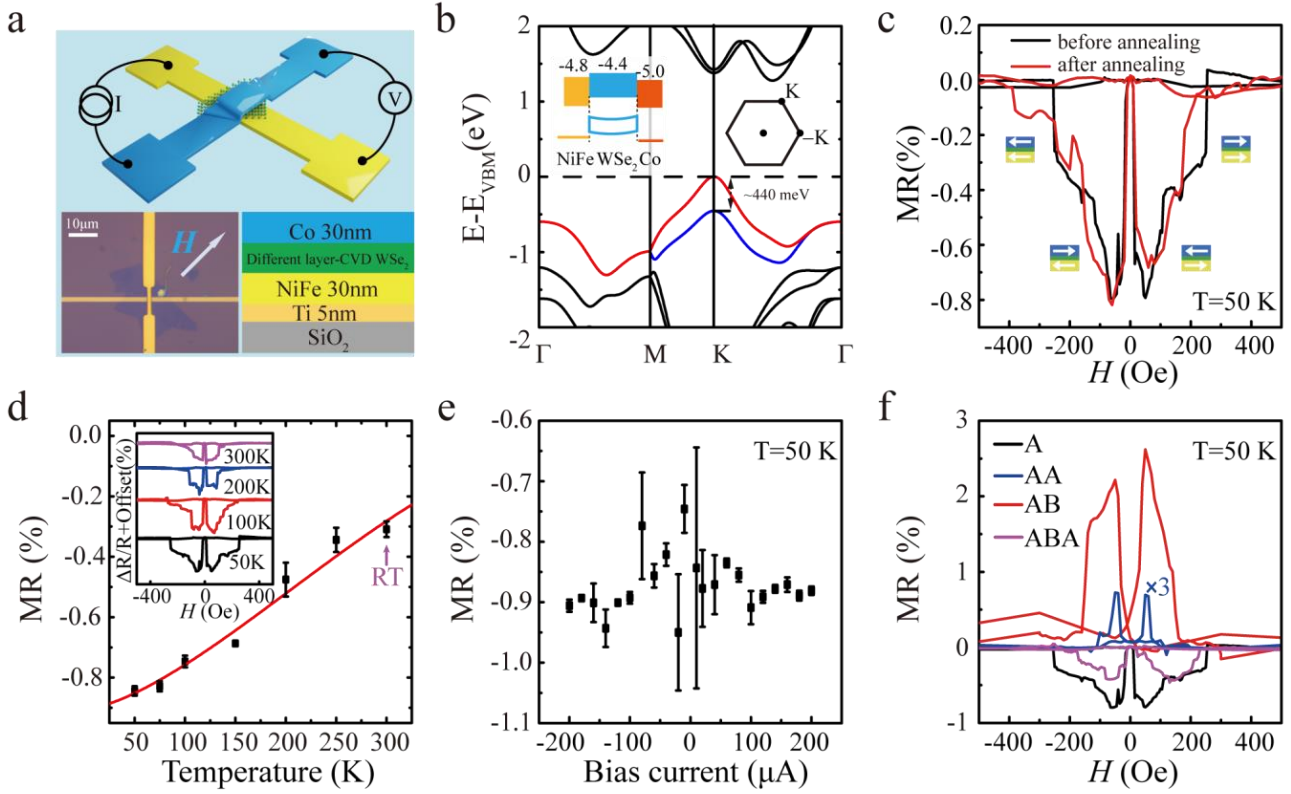

**FIG. 2.** Nontrivial MR of the WSe$_2$ vertical spin valve. (a) Schematic structure and optical image of the WSe$_2$ vertical spin valve. (b) Band structure of monolayer WSe$_2$ using first-principles calculation. The spitting sub-bands of out-of-plane spin are indicated by the red and blue lines. Inset: Left: Band alignments of WSe$_2$ vertical spin valve with NiFe and Co electrodes. Right: The diagram of Brillouin Zone where K and -K have opposite spin splittings. (c) Negative MR of monolayer WSe$_2$. The semi-parallel and semi-antiparallel magnetization alignments between two FM electrodes are marked. The black (red) line shows the MR before (after) thermal annealing treatment. (d)&(e) The temperature (d) and bias (e) dependence of negative MR in monolayer WSe$_2$. The red line in (d) is the fitting result according to Bloch's law. (f) The layer number and stacking phase dependence of MR. The significant oscillation is reproducible in various devices. The measurement temperature of (c), (e) and (f) was 50 K.

In 2D material vertical junction devices, physical effects, such as the interfacial effect (hybridization, magnetic proximity, or Rashba effect), spin filtering effect, interlayer exchange coupling, organic MR, and spin-polarized resonant tunneling, can contribute to abundant MR effects[35-41]. After excluding these possible origins, we suggest that the observations in our devices were caused by the giant valley-Zeeman SOF in WSe$_2$. (See S6–S8 for discussion).

A critical feature of the valley-Zeeman SOF in WSe$_2$ is the tunability of this field by either the layer number or stacking phase. Accordingly, we found an oscillation in the sign of MR and a reduction in the amplitude of the MR value when the number of WSe$_2$ layers is increased, as shown in FIG. 2(f), which are reproducible in a dozen of devices (see S9 for reproducibility). Particularly, the positive MR in bilayer WSe$_2$ decreases from +2.5% in AB stacking to +0.3% in AA stacking.

To confirm the valley-Zeeman SOF-induced spin dynamics theoretically, the analysis of the quantum transmission near the K-valley is performed. Considering the valley-Zeeman SOF by the SOC term $\boldsymbol{L \cdot S}$, the $k \cdot p$ Hamiltonian of monolayer WSe$_2$ near the K point can be expressed as[21]:

$$H = v_f(\tau k_x\sigma_x + k_y\sigma_y) + \frac{\Delta}{2}\sigma_z - \Delta_{so}\tau\frac{\sigma_z-1}{2}S_z - \frac{\hbar^2\partial^2}{2m\partial z^2} - V_G, \quad (1)$$

where $\sigma_i$ is the Pauli matrix in the orbital space, $S_z$ is the spin matrix; $\tau = \pm1$ is the index for the ±K valley; $2\Delta_{so}$ is the strength of SOC, $\Delta$ is the band gap, and $v_f$ is the velocity; the second last term is the kinetic energy along the z-axis within monolayer WSe$_2$; and $V_G$ is the gating potential used to adjust the Fermi level of WSe$_2$. The WSe$_2$ parameters listed in Table 1 were obtained by fitting the first-principles calculations (see S10 for details). For the cases of bilayer WSe$_2$, the Hamiltonian of the scattering region in Eq. (1) can be considered in layers, with a hopping term $\frac{\hbar^2}{2m_i}\partial^2_z$ between layers, where $m_i$ =1.6$m_0$ is the effective mass between layers by fitting to the band of bulk WSe$_2$ along c-axis. For AA stacking, two adjacent layers have the same sign of SOC at K point, while the sign is opposite for AB stacking (see Fig. S9 for bands of AA and AB stacking).

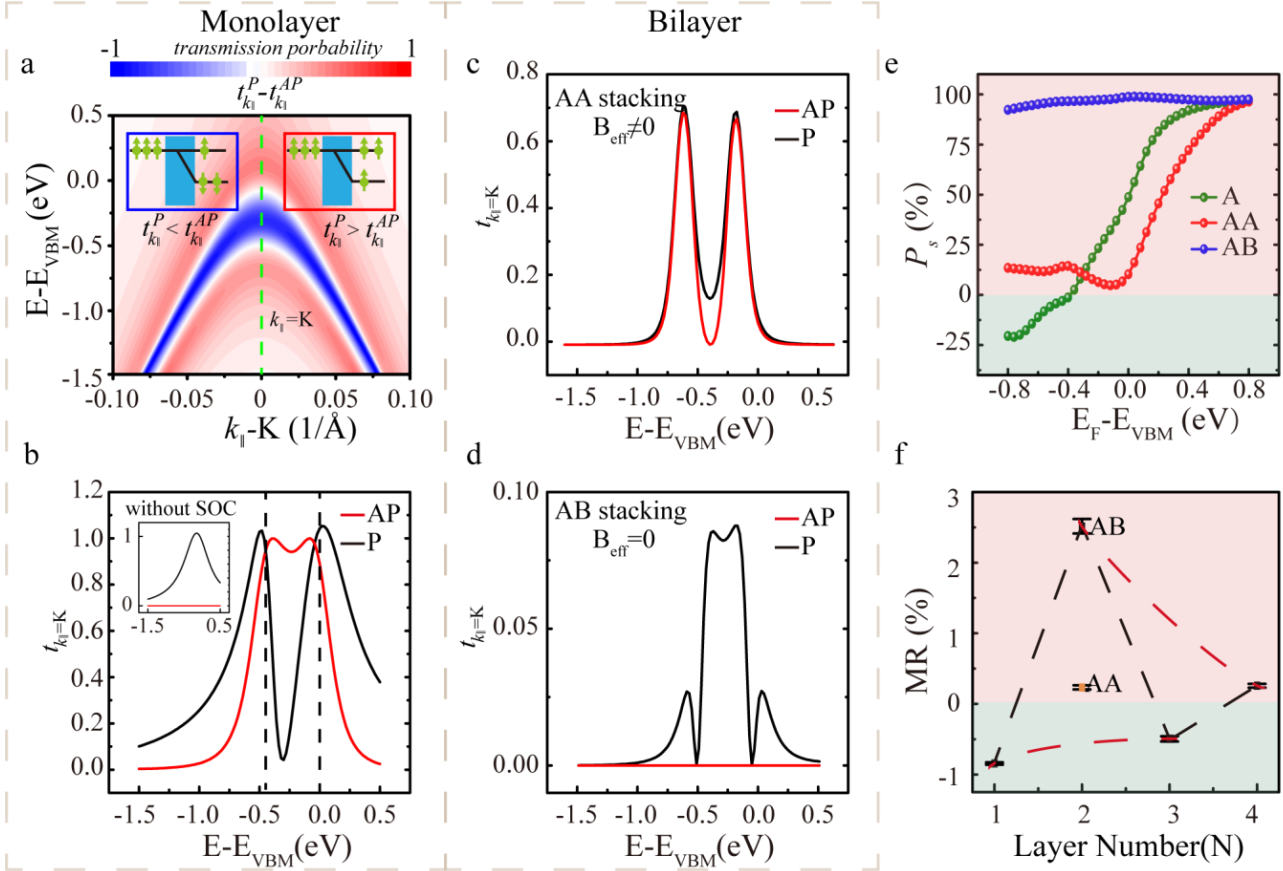

**FIG 3** Calculations of spin-resolved quantum transmission in a WSe$_2$ vertical spin valve. (a) The $k_\parallel$-resolved differential transmission spectrum between P and AP states of the device with monolayer WSe$_2$, where $t_{k_\parallel}^{P(AP)}$ is the transmission probability for (anti-) parallel states. Inset: Schematic diagram of spin flipping during scattering. (b) The transmission probability at the K point of monolayer WSe$_2$, where the dashed lines represent the location of the sub-band. Inset: The situation without spin-orbital coupling ($\Delta_{so}=0$). (c), (d) The transmission probability at the K point of the bilayer WSe$_2$ vertical spin valve with AA and AB stacking, respectively. (e) The Fermi level dependence of spin polarization of carriers scattering through monolayer and bilayer WSe$_2$, where the spin polarization is calculated from the integral of $t_{k_\parallel}^{P}$ and $t_{k_\parallel}^{AP}$ at corresponding Fermi level. The green (bule and red) line represents the monolayer (bilayer with AB and AA stacking) case. (f) The fitting results (orange dashed line) for layer dependence of MR according to the modified Julliere's model. The error bars in (f) result from the averaging of measurements.

We then built a two-terminal device with tight-binding model to perform the transmission calculations (see FIG. S8), where two semi-infinite regions along the z-direction with 100% spin polarization oriented along the x-axis were used to model the leads, and the Fermi level was adjusted at the VBM of WSe$_2$.

**Table 1** Fitting result from band structure of first-principles calculations, where $m_0$ is the electron mass.

| $v_f$ | $\Delta_{so}$ | $\Delta$ | $m$ |
|---|---|---|---|
| 22.3 | 0.22 eV | 1.6 eV | $-0.36m_0$ |

The $k_\parallel$-resolved transmission probability spectrum was calculated with the scattering theory by the Kwant package[42]. Because the transport through atom-thick WSe$_2$ is ballistic, each $k_\parallel$ state contributes to the transmission independently, then the transmission coefficient $T$ can be calculated according to the Buttiker-Landauer formula:

$$T_{P(AP)}(E) = \sum_{k_{\parallel}} \frac{e^2}{h} t_{k_{\parallel}}^{P(AP)}(E), \quad (2)$$

where $t_{k_{\parallel}}^{P}$ and $t_{k_{\parallel}}^{AP}$ are the transmission probabilities of the P and AP states at each $k_{\parallel}$ point, respectively. The spectrum of the differential transmission probability between the P and AP states ($t_{k_{\parallel}}^{P} - t_{k_{\parallel}}^{AP}$) is plotted in FIG. 3(a). Nontrivially, $t_{k_{\parallel}}^{P}$ is smaller than $t_{k_{\parallel}}^{AP}$ at the region between two sub-bands of the K valley. The appearance of $t_{k_{\parallel}}^{AP} > t_{k_{\parallel}}^{P}$ contributes to the negative MR of WSe$_2$ spin valve.

To better understand the unconventional MR under valley-Zeeman SOF, the transmission channel at the K-point is taken as an example, corresponding to the green dash line shown in FIG. 3(a). In the absence of SOC term in Eq. (1), the spin of incident carriers cannot be deflected during the scattering process. Therefore, the incident carriers will be reflected completely by the bottom electrode with opposite spin polarization, that is, $t_{k_{\parallel}}^{AP} = 0$ (inset of FIG. 3(b)), corresponding to an ordinary positive MR. However, in the presence of SOC term, both the P and AP states contribute to the transmission coefficients, as shown in FIG. 3(b), which indicates the spin deflection and even flipping under valley-Zeeman SOF.

The valley-Zeeman SOF results in an oscillating sign of MR with the layer number. According to the transmission probability at K channel of the bilayer WSe$_2$ (FIGs. 3(c) and (d)), in the case of AB stacking, the transmission appears only in the P state, which means no deflection of spin. In the AA stacking WSe$_2$, although transmission appears in both the P and AP states, $t_{k_{\parallel}}^{P}$ is always larger than $t_{k_{\parallel}}^{AP}$. Then, we calculated the transmission coefficient $T_{P(AP)}$ according to the Buttiker-Landauer formula (Eq. (2), also see FIG. S10 for the transmission spectrum of bilayer WSe$_2$) and plotted the spin polarization $P_s = \frac{T_P - T_{AP}}{T_P + T_{AP}}$ of incident carriers transmitted through monolayer and bilayer WSe$_2$ in FIG. 3(e). According to the Julliere's model, the MR can be described as $MR = \frac{2P_1 P_s P_2}{1 + P_1 P_s P_2}$, where $P_1$ and $P_2$ represent the interfacial spin polarization of FM contacts. At $E_F - E_{VBM} \approx -0.5$ eV, $P_s$ is negative for the case of monolayer, while positive for bilayer WSe$_2$ with both AB and AA stacking, which is consistent with the oscillated sign of MR in experiment. We also note that the spin-degenerate energy band at the $\Gamma$ point in multilayer WSe$_2$ can contribute to positive MR at $E-E_F = -0.5$ eV. While for monolayer WSe$_2$ only the bands of the two valleys appear around $E-E_F = -0.5$ eV and thus the contribution from $\Gamma$ point does not affect our conclusion about the layer-dependent oscillation of MR.

The valley-Zeeman SOF also plays a crucial role in the quenching of the MR amplitude with the layer number. As shown in FIG. 3(e), the amplitude of $P_s$ decreases to 10% at $E_F - E_{VBM} \approx -0.5$ eV in both case of monolayer and AA stacking WSe$_2$, which can be understood by the spin deflection under valley-Zeeman SOF (both $t_{k_{\parallel}}^{P}$ and $t_{k_{\parallel}}^{AP}$ are nonzero). Non-strictly but effectively, we introduce the exponential form to describe the spin dissipation in our device. For the ordinary case, that is, the decay of spin polarization during the carrier propagation[43], the spin dissipations can be considered as the term $\exp(-\beta_0 L)$ with dissipation coefficient $\beta_0$ and transport length $L$. For the additional spin dissipation by valley-Zeeman SOF, its contribution to the spin dissipations can be considered as $|P_s| = \exp(-\beta_s L)$ with an effective spin dissipation coefficient $\beta_s$. The quenching of MR amplitude is evaluated using the modified Julliere's model[44]: $MR = 2P_1 P_2 \exp(-\beta L)/(1 + P_1 P_2 \exp(-\beta L))$, where $\beta$ is total spin dissipation coefficient. The transport length $L$ is estimated as $L_N = N \cdot d$ for WSe$_2$ with different layers ($N$), where $d = 0.8$ nm is the thickness of single-layer WSe$_2$. Accordingly, the spin dissipation is $\exp(-(\beta_0 + \beta_s)d)$ for the monolayer, $\exp(-2\beta_0 d)$ for the bilayer with AB stacking, $\exp(-2(\beta_0 + \beta_s)d)$ for the bilayer with AA stacking, $\exp(-(3\beta_0 + \beta_s)d)$ for the trilayers with ABA stacking, and $\exp(-4\beta_0 d)$ for the quadlayers with ABAB stacking. By regression analysis and curve fitting (see FIG. S11), the quenching behavior of the MR can be successfully described with $\beta_s = 2.41\beta_0$, as shown FIG 3(f).

Therefore, the intrinsic physical mechanism of the unconventional MR in the WSe$_2$ vertical spin valve can be concluded as valley-Zeeman SOF-induced spin dynamics during the scattering process. Although the transport needs to be described by quantum theory because of the atom-thick transport channel in the vertical device structure, we can still obtain an intuitive understanding by virtue of the semiclassical treatment, as a correspondence with the spin precession under Rashba-type SOF[11, 12]. To validate the semiclassical treatment, we studied the propagation properties of electronic wave packet through thicker WSe$_2$ with a small Zeeman-type splitting $\Delta_{so} = 0.05$ eV. The manifestation that in-plane spin rotates with a certain frequency is conformed to the precessional picture (see S12 for details). The precession frequency with small $\Delta_{so}$ implies that the spin precession angle per layer can be roughly estimated as $1.2\pi$ from the Zeeman splitting of monolayer WSe$_2$.

Furthermore, the magnitude of the valley-Zeeman SOF $B_{eff}$ in monolayer WSe$_2$ can be evaluated from $\Delta_{so} = g\mu_B B_{eff}$ with a reasonable Landé g-factor, where $\mu_B \approx 0.0579$ meV/T is the Bohr magneton. The effective g-factor of the WSe$_2$ band is understood including three contributions [45, 46], namely the spin ($g_s = 1$, in our situation), the tungsten d-orbital ($g_o = 2$ for the valence band) and the valley associated with Berry curvature ($g_v$).

The last one can be extracted from a massive Dirac fermion model [45] with $g_v = m_0/m$, where $m$ and $m_0$ are the intrinsic and effective mass of electrons, respectively. For the upper valence band of $WSe_2$, $m$ is expected as 0.36 $m_0$[25]. The overall g-factor is the sum of these three parts, i.e., $g \approx 5.8$, according to the results of excitonic resonances [46, 47]. Therefore, the valley-Zeeman SOF at the VBM is estimated as $B_{eff}^M \approx 650$ T, which is a huge value not achievable by magnetic fields produced in laboratories.

By contrast, the coherent spin precession of the Rashba spin-FET requires a long ballistic transport channel to flip spin, which requires low-temperature and high-quality heterostructures. Owing to the large value of the effective magnetic field, the valley-Zeeman SOF can drive complete spin-flipping even in a vertical device of atomic-thick $WSe_2$, within the region of room-temperature ballistic quantum transport.

Importantly, the location of Fermi level is crucial for observing the nontrivial MR in TMDs, as the valley-Zeeman SOF can be "screened" when the Fermi level is adjusted away from the top of the valence band[48]. We also performed similar measurement in devices with $MoS_2$. In this case, the Fermi level is located at the bottom of the conduction band of $MoS_2$, where the SOC splitting is quite weak. We found that the MR is always positive and increases with the number of layers (see FIG. S4 and Ref[49]), which can be explained by the spin-filtering effect[34]. What is more, as shown in FIG. 3(e), when we adjusted the $V_g$ term in Eq. (1) during the transmission calculation, the sign of $P_s$ in the case of monolayer $WSe_2$ can be changed according to the location of Fermi level. The gate-tunable valley-Zeeman SOF provides a strategy for the electrical manipulation of spin transport. Compared to the electrical control of carrier diffusion in micron channel with proximity effect of valley-Zeeman SOF [18, 50], our results indicate that the spin current can be controlled within vertical atom-thick channel by giant valley-Zeeman SOF of $WSe_2$, which can further benefit the high-density integration. Correspondingly, a valley-Zeeman SOF logic device can be proposed (see Fig. S13), which utilizes the gate voltage to switch the effective field and thus spin flipping to implement logic at room temperature. Furthermore, the recent demonstrated interfacial ferroelectricity provides a new strategy of electric tunability in rhombohedral-stacked TMDs[51, 52], which may induce a control knob for the valley-Zeeman SOF and highlights the great potential of spin manipulation.

## CONCLUSION

In conclusion, we observed valley-Zeeman SOF-induced spin dynamics through atom-thick $WSe_2$. This physical effect has been demonstrated as an unconventional MR effect in a $WSe_2$ vertical spin valve, where the valley-Zeeman SOF varies with the $WSe_2$ layer number and stacking phase. Supported by quantum transport calculations, we further clarified the spin-flipping mechanism driven by the valley-Zeeman SOF in two-dimensional $WSe_2$. This mechanism provides a new avenue for the control of spin transport even at room temperature, which is promising for realizing high-density and ultralow-power spintronic devices.

**Supplementary Material**
See the supplementary material for details on sample fabrication and characterizations, the discussion on band alignment, the fitting of temperature dependence of MR, the analysis of anisotropic MR of ferromagnetic layers, additional experimental results of control devices, the discussion on negative MR in vertical spin valves, the oscillation of MR in various devices, the details of quantum transmission calculation, the discussion and fitting of spin dissipation, the calculation of electronic wave packet, and the proposed spin logic device based Zeeman SOF.

## ACKNOWLEDGMENTS

We acknowledge helpful discussions with Albert Fert and Yong Xu. We thank Lei Liu's contributions to our first-principles calculations. This work was supported by the National Natural Science Foundation of China (No. 51602013, 11904014, 11804016, 61627813 and 62174010), Young Elite Scientists Sponsorship Program by China Association for Science and Technology (CAST) (No. 2018QNRC001), the International Collaboration 111 Project (No. B16001), Beijing Natural Science Foundation(No. 4222070), the Fundamental Research Funds for the Central Universities of China, and the Beijing Advanced Innovation Centre for Big Data and Brain Computing (BDBC).

X.W., Wei Y., Wang Y. and Y. C. contributed equally to this work.